\input epsf

\newfam\scrfam
\batchmode\font\tenscr=rsfs10 \errorstopmode
\ifx\tenscr\nullfont
        \message{rsfs script font not available. Replacing with calligraphic.}
        
\else   
        \font\sevenscr=rsfs7
        \font\fivescr=rsfs5
        \skewchar\tenscr='177 \skewchar\sevenscr='177 \skewchar\fivescr='177
        \textfont\scrfam=\tenscr \scriptfont\scrfam=\sevenscr
        \scriptscriptfont\scrfam=\fivescr

\fi
\catcode`\@=11
\newfam\frakfam
\batchmode\font\tenfrak=eufm10 \errorstopmode
\ifx\tenfrak\nullfont
        \message{eufm font not available. Replacing with italic.}
        
\else
	
	\font\sevenfrak=eufm7 \font\fivefrak=eufm5
        
	\textfont\frakfam=\tenfrak
	\scriptfont\frakfam=\sevenfrak \scriptscriptfont\frakfam=\fivefrak
	
\fi
\catcode`\@=\active
\newfam\msbfam
\batchmode\font\twelvemsb=msbm10 scaled\magstep1 \errorstopmode
\ifx\twelvemsb\nullfont\def\Bbb{\bf}
        
	\font\eightbbb=cmb10 at 8pt
	\message{Blackboard bold not available. Replacing with boldface.}
\else   \catcode`\@=11
        \font\tenmsb=msbm10 \font\sevenmsb=msbm7 \font\fivemsb=msbm5
        \textfont\msbfam=\tenmsb
        \scriptfont\msbfam=\sevenmsb \scriptscriptfont\msbfam=\fivemsb
        \def\Bbb{\relax\expandafter\Bbb@}
        \def\Bbb@#1{{\Bbb@@{#1}}}
        \def\Bbb@@#1{\fam\msbfam\relax#1}
        \catcode`\@=\active
	
	\font\eightbbb=msbm8
\fi
        \font\fivemi=cmmi5
        \font\sixmi=cmmi6
        \font\eightrm=cmr8              \def\xrm{\eightrm}
        \font\eightbf=cmbx8             \def\xbf{\eightbf}
        \font\eightit=cmti10 at 8pt     \def\xit{\eightit}
                
        \font\eighttt=cmtt8             
        \font\eightcp=cmcsc8
        \font\eighti=cmmi8              \def\xold{\eighti}
        \font\eightmi=cmmi8
        \font\eightib=cmmib8             \def\xbold{\eightib}
        \font\teni=cmmi10               \def\old{\teni}
        \font\tencp=cmcsc10

        \font\twelvecp=cmcsc10 scaled\magstep1
        
        \font\sixrm=cmr6
        \font\fiverm=cmr5

        \font\eightsy=cmsy8
        \font\sixsy=cmsy6
        \font\eightsl=cmsl8
        \font\sixbf=cmbx6

	 at10pt	
	\font\twelvehelvbold=phvb at12pt
	 at14pt
	\font\sixteenhelvbold=phvb at16pt

\def\noblackbox{\overfullrule=0pt}
\noblackbox

\def\eightpoint{
\def\rm{\fam0\eightrm}
\textfont0=\eightrm \scriptfont0=\sixrm \scriptscriptfont0=\fiverm
\textfont1=\eightmi  \scriptfont1=\sixmi  \scriptscriptfont1=\fivemi
\textfont2=\eightsy \scriptfont2=\sixsy \scriptscriptfont2=\fivesy
\textfont3=\tenex   \scriptfont3=\tenex \scriptscriptfont3=\tenex
\textfont\itfam=\eightit \def\it{\fam\itfam\eightit}
\textfont\slfam=\eightsl \def\sl{\fam\slfam\eightsl}
\textfont\ttfam=\eighttt \def\tt{\fam\ttfam\eighttt}
\textfont\bffam=\eightbf \scriptfont\bffam=\sixbf 
                         \scriptscriptfont\bffam=\fivebf
                         \def\bf{\fam\bffam\eightbf}
\normalbaselineskip=10pt}

\newtoks\headtext
\headline={\ifnum\pageno=1\hfill\else
	\ifodd\pageno{\eightcp\the\headtext}{ }\dotfill{ }{\old\folio}
	\else{\old\folio}{ }\dotfill{ }{\eightcp\the\headtext}\fi
	\fi}
\def\makeheadline{\vbox to 0pt{\vss\noindent\the\headline\break
\hbox to\hsize{\hfill}}
        \vskip2\baselineskip}
\newcount\infootnote
\infootnote=0
\newcount\footnotecount
\footnotecount=1
\def\foot#1{\infootnote=1
\footnote{${}^{\the\footnotecount}$}{\vtop{\baselineskip=.75\baselineskip
\advance\hsize by
-\parindent{\eightpoint\rm\hskip-\parindent
#1}\hfill\vskip\parskip}}\infootnote=0\global\advance\footnotecount by
1}
\newcount\refcount
\refcount=1
\newwrite\refwrite
\def\oldsize{\ifnum\infootnote=1\xold\else\old\fi}
\def\ref#1#2{
	\def#1{{{\oldsize\the\refcount}}\ifnum\the\refcount=1\immediate\openout\refwrite=\jobname.refs\fi\immediate\write\refwrite{\item{[{\xold\the\refcount}]} 
	#2\hfill\par\vskip-2pt}\xdef#1{{\noexpand\oldsize\the\refcount}}\global\advance\refcount by 1}
	}
\def\refout{\eightpoint\catcode`\@=11
        \xrm\immediate\closeout\refwrite
        \vskip2\baselineskip
        {\noindent\twelvecp References}\hfill\vskip\baselineskip
        \baselineskip=.75\baselineskip
        \input\jobname.refs
        \baselineskip=4\baselineskip \divide\baselineskip by 3
        \catcode`\@=\active\rm}

\def\skipref#1{\hbox to15pt{\phantom{#1}\hfill}\hskip-15pt}

\def\hepth#1{\href{http://xxx.lanl.gov/abs/hep-th/#1}{arXiv:\allowbreak
hep-th\slash{\xold#1}}}

\def\arxiv#1#2{\href{http://arxiv.org/abs/#1.#2}{arXiv:\allowbreak
{\xold#1}.{\xold#2}}} 
\def\jhep#1#2#3#4{\href{http://jhep.sissa.it/stdsearch?paper=#2\%28#3\%29#4}{J. High Energy Phys. {\xbold #1#2} ({\xold#3}) {\xold#4}}}

\def\CQG#1#2#3{Class. Quantum Grav. {\xbold#1} ({\xold#2}) {\xold#3}}

\def\JPA#1#2#3{J. Phys. {\xbf A}{\xbold#1} ({\xold#2}) {\xold#3}}
\def\LMP#1#2#3{Lett. Math. Phys. {\xbold#1} ({\xold#2}) {\xold#3}}
\def\MPLA#1#2#3{Mod. Phys. Lett. {\xbf A}{\xbold#1} ({\xold#2}) {\xold#3}}

\def\NPB#1#2#3{Nucl. Phys. {\xbf B}{\xbold#1} ({\xold#2}) {\xold#3}}

\def\PLB#1#2#3{Phys. Lett. {\xbf B}{\xbold#1} ({\xold#2}) {\xold#3}}
\def\PR#1#2#3{Phys. Rept. {\xbold#1} ({\xold#2}) {\xold#3}}
\def\PRD#1#2#3{Phys. Rev. {\xbf D}{\xbold#1} ({\xold#2}) {\xold#3}}
\def\PRL#1#2#3{Phys. Rev. Lett. {\xbold#1} ({\xold#2}) {\xold#3}}

\newcount\sectioncount
\sectioncount=0
\def\section#1#2{\global\eqcount=0
	\global\subsectioncount=0
        \global\advance\sectioncount by 1
	\ifnum\sectioncount>1
	        \vskip2\baselineskip
	\fi
\line{\twelvecp\the\sectioncount. #2\hfill}
       \vskip.5\baselineskip\noindent
        \xdef#1{{\old\the\sectioncount}}}
\newcount\subsectioncount
\def\subsection#1#2{\global\advance\subsectioncount by 1
\vskip.75\baselineskip\noindent\line{\tencp\the\sectioncount.\the\subsectioncount. #2\hfill}\nobreak\vskip.4\baselineskip\nobreak\noindent\xdef#1{{\old\the\sectioncount}.{\old\the\subsectioncount}}}
\def\immediatesubsection#1#2{\global\advance\subsectioncount by 1
\vskip-\baselineskip\noindent
\line{\tencp\the\sectioncount.\the\subsectioncount. #2\hfill}
	\vskip.5\baselineskip\noindent
	\xdef#1{{\old\the\sectioncount}.{\old\the\subsectioncount}}}
\newcount\subsubsectioncount
\def\subsubsection#1#2{\global\advance\subsubsectioncount by 1
\vskip.75\baselineskip\noindent\line{\tencp\the\sectioncount.\the\subsectioncount.\the\subsubsectioncount. #2\hfill}\nobreak\vskip.4\baselineskip\nobreak\noindent\xdef#1{{\old\the\sectioncount}.{\old\the\subsectioncount}.{\old\the\subsubsectioncount}}}
\newcount\appendixcount
\appendixcount=0
\def\appendix#1{\global\eqcount=0
        \global\advance\appendixcount by 1
        \vskip2\baselineskip\noindent
        \ifnum\the\appendixcount=1
        \hbox{\twelvecp Appendix A: #1\hfill}\vskip\baselineskip\noindent\fi
    \ifnum\the\appendixcount=2
        \hbox{\twelvecp Appendix B: #1\hfill}\vskip\baselineskip\noindent\fi
    \ifnum\the\appendixcount=3
        \hbox{\twelvecp Appendix C: #1\hfill}\vskip\baselineskip\noindent\fi}
\def\acknowledgements{\vskip2\baselineskip\noindent
        \underbar{\it Acknowledgements:}\ }
\newcount\eqcount
\eqcount=0
\def\Eqn#1{\global\advance\eqcount by 1
\ifnum\the\sectioncount=0
	\xdef#1{{\noexpand\oldsize\the\eqcount}}
	\eqno({\oldstyle\the\eqcount})
\else
        \ifnum\the\appendixcount=0
\xdef#1{{\noexpand\oldsize\the\sectioncount}.{\noexpand\oldsize\the\eqcount}}
                \eqno({\oldstyle\the\sectioncount}.{\oldstyle\the\eqcount})\fi
        \ifnum\the\appendixcount=1
	        \xdef#1{{\noexpand\oldstyle A}.{\noexpand\oldstyle\the\eqcount}}
                \eqno({\oldstyle A}.{\oldstyle\the\eqcount})\fi
        \ifnum\the\appendixcount=2
	        \xdef#1{{\noexpand\oldstyle B}.{\noexpand\oldstyle\the\eqcount}}
                \eqno({\oldstyle B}.{\oldstyle\the\eqcount})\fi
        \ifnum\the\appendixcount=3
	        \xdef#1{{\noexpand\oldstyle C}.{\noexpand\oldstyle\the\eqcount}}
                \eqno({\oldstyle C}.{\oldstyle\the\eqcount})\fi
\fi}
\def\eqn{\global\advance\eqcount by 1
\ifnum\the\sectioncount=0
	\eqno({\oldstyle\the\eqcount})
\else
        \ifnum\the\appendixcount=0
                \eqno({\oldstyle\the\sectioncount}.{\oldstyle\the\eqcount})\fi
        \ifnum\the\appendixcount=1
                \eqno({\oldstyle A}.{\oldstyle\the\eqcount})\fi
        \ifnum\the\appendixcount=2
                \eqno({\oldstyle B}.{\oldstyle\the\eqcount})\fi
        \ifnum\the\appendixcount=3
                \eqno({\oldstyle C}.{\oldstyle\the\eqcount})\fi
\fi}
\def\multi{\global\advance\eqcount by 1}
\def\multieqn#1{({\oldstyle\the\sectioncount}.{\oldstyle\the\eqcount}\hbox{#1})}
\def\multiEqn#1#2{\xdef#1{{\oldstyle\the\sectioncount}.{\old\the\eqcount}#2}
        ({\oldstyle\the\sectioncount}.{\oldstyle\the\eqcount}\hbox{#2})}
\def\multiEqnAll#1{\xdef#1{{\oldstyle\the\sectioncount}.{\old\the\eqcount}}}
\newcount\tablecount
\tablecount=0
\def\Table#1#2{\global\advance\tablecount by 1
       \xdef#1{\the\tablecount}
       \vskip2\parskip
       \centerline{\it Table \the\tablecount: #2}
       \vskip2\parskip}
\newtoks\url
\def\Href#1#2{\catcode`\#=12\url={#1}\catcode`\#=\active#2}
\def\href#1#2{{#2}}

\parskip=3.5pt plus .3pt minus .3pt
\baselineskip=14pt plus .1pt minus .05pt
\lineskip=.5pt plus .05pt minus .05pt
\lineskiplimit=.5pt
\abovedisplayskip=18pt plus 4pt minus 2pt
\belowdisplayskip=\abovedisplayskip
\hsize=14cm
\vsize=19cm
\hoffset=1.5cm
\voffset=1.8cm
\frenchspacing
\footline={}
\raggedbottom

\newskip\origparindent
\origparindent=\parindent

\def\*{\partial}
\def\punkt{\,\,.}
\def\komma{\,\,,}

\def\={\!=\!}
\def\small#1{{\hbox{$#1$}}}

\def\fraction#1{\small{1\over#1}}
\def\fr{\fraction}
\def\Fraction#1#2{\small{#1\over#2}}
\def\Fr{\Fraction}

\def\eg{{\it e.g.}}

\def\ie{{\it i.e.}}

\def\nlni{\hfill\break}

\def\a{\alpha}
\def\b{\beta}
\def\d{\delta}
\def\e{\varepsilon}
\def\g{\gamma}

\def\s{\sigma}

\def\L{\Lambda}

\def\ra{\rightarrow}

\def\ra{\rightarrow}

\catcode`@=11                                   
\catcode`\|=12                                  
\catcode`\&=4                                   

\newcount\ncols         \ncols=\z@              
\newcount\nrows         \nrows=\z@              
\newcount\curcol        \curcol=\z@             
     
\newdimen\thinsize      \thinsize=0.6pt         
\newdimen\thicksize     \thicksize=1.5pt        

\newif\iftableinfo      \tableinfotrue          
\newif\ifcentertables   \centertablestrue       
%
%
     
\let\plaincr=\cr                        
\let\plainspan=\span                    
\let\plaintab=&                         
\let\lparen=(                           
\let\NX=\noexpand                       

     
\def\ruledtable{\relax                          
    \@BeginRuledTable                           
    \@RuledTable}


\def\@BeginRuledTable{
   \ncols=0\nrows=0                             
   \begingroup                                  
    \offinterlineskip                           
    \def~{\phantom{0}}
    \def\span{\plainspan\omit\relax\colcount\plainspan}
    \let\cr=\crrule                             
    \let\CR=\crthick                            
    \let\nr=\crnorule                           
    \let\|=\Vb                                  
%
%
    \ifx\tablestrut\undefined\relax             
    \else\let\tstrut=\tablestrut\fi             
    \catcode`\|=13 \catcode`\&=13\relax         
    \TableActive                                
    \curcol=1                                   
%
%
    \ifdim\tablewidth>-\maxdimen\relax          %
      \edef\@Halign{\NX\halign to \NX\tablewidth\NX\bgroup\TablePreamble}%
      \tabskip=0pt plus 1fil                    
    \else                                       %
      \edef\@Halign{\NX\halign\NX\bgroup\TablePreamble}%
      \tabskip=0pt                              
    \fi                                         %
%
%
    \ifcentertables                             
       \ifhmode\vskip 0pt\fi                    
       \line\bgroup\hss                         
    \else\hbox\bgroup                           
    \fi}


\long\def\@RuledTable#1\endruledtable{
   \vrule width\thicksize                       
     \vbox{\@Halign                             
       \thickrule                               
       #1\relax                                 
       \tstrut                                  
       \plaincr\thickrule                       
     \egroup}
   \vrule width\thicksize                       
   \ifcentertables\hss\fi\egroup                
  \endgroup                                     
  \global\tablewidth=-\maxdimen                 
  \iftableinfo                                  
      \immediate\write16{[Nrows=\the\nrows, Ncols=\the\ncols]}%
   \fi}
     

\def\TablePreamble{
   \linecount                           
   \TableItem{####}
   \plaintab\plaintab                   
   \TableItem{####}
   \plaincr}


\def\@TableItem#1{
   \hfil\tablespace                             
   #1\relax                                     
   \tablespace\hfil                             
    }%

\def\@tableright#1{
   \hfil\tablespace\relax               
   #1\relax                             
   \tablespace\relax}

\def\@tableleft#1{
   \tablespace\relax                    
   #1\relax                             
   \tablespace\hfil}

\let\TableItem=\@TableItem              
     
\def\RightJustifyTables{\let\TableItem=\@tableright}
\def\LeftJustifyTables{\let\TableItem=\@tableleft}
\def\NoJustifyTables{\let\TableItem=\@TableItem}

\def\LooseTables{\let\tablespace=\quad}
\def\TightTables{\let\tablespace=\space}
\LooseTables                                    

%

\newdimen\tablewidth    \tablewidth=-\maxdimen  


\def\setRuledStrut{
   \dimen@=\baselineskip                        
   \advance\dimen@ by-\normalbaselineskip       
   \ifdim\dimen@<.5ex \dimen@=.5ex\fi           
   \setbox0=\hbox{\lparen}
   \dimen1=\dimen@ \advance\dimen1 by \ht0      
   \dimen2=\dimen@ \advance\dimen2 by \dp0      
   \def\tstrut{\vrule height\dimen1 depth\dimen2 width\z@}%
   }%

\def\tstrut{\vrule height 3.1ex depth 1.2ex width 0pt}


\def\bigitem#1{
   \setbox0=\hbox{#1}
   \dimen1 =\ht0 \dimen2 =\dp0                  
   \dimen@ =\baselines@ve                       
   \advance\dimen@ by-\normalbaselineskip       
   \ifdim\dimen@<.25ex \dimen@=.25ex\fi         
   \advance\dimen1 by \dimen@                   
   \advance\dimen2 by \dimen@                   
   \vrule height\dimen1 depth\dimen2 width\z@   
   \copy0}

     
%

     
\def\nextcolumn#1{
   \plaintab\omit#1\relax\colcount              
   \plaintab}
     
\def\tab{
   \nextcolumn{\relax}}


\def\vb{
   \nextcolumn{\vrule width\thinsize}}

\def\Vb{
   \nextcolumn{\vrule width\thicksize}}


     
{\catcode`\|=13 \let|0
 \catcode`\&=13 \let&0
 \gdef\TableActive{\let|=\vb \let&=\tab}%
}


\def\crrule{\relax                      
   \tstrut                              
   \plaincr\tablerule                   
  }%

\def\crthick{\relax                     
   \tstrut                              
   \plaincr\thickrule                   
  }%
     
\def\crnorule{\relax                    
   \tstrut                              
   \plaincr                             
   }%
   

     
\def\tablerule{\noalign{\hrule height\thinsize depth 0pt}}%
\def\thickrule{\noalign{\hrule height\thicksize depth 0pt}}%


%
%
%
     

\def\linecount{\relax\global\ncols=\curcol      
   \global\curcol=1                             
   \global\advance\nrows by 1\relax}
     
\def\colcount{\relax                            %
   \global\advance\curcol by 1\relax}


\newdimen\parasize      \parasize=4in           

%

%

\def\begintable{\relax                          
    \@BeginRuledTable                           
    \@begintable}

\long\def\@begintable#1\endtable{
   \@RuledTable#1\endruledtable}


\catcode`@=12                                   




\def\L{\Lambda}

\def\ol{\overline}

\def\textfrac#1#2{\raise .45ex\hbox{\the\scriptfont0 #1}\nobreak\hskip-1pt/\hskip-1pt\hbox{\the\scriptfont0 #2}}


\def\frac{\Fr}

\def\mathbb{\Bbb}



\ref\CederwallUfoldbranes{M. Cederwall, {\xit ``M-branes on U-folds''},
in proceedings of 7th International Workshop ``Supersymmetries and
Quantum Symmetries'' Dubna, 2007 [\arxiv{0712}{4287}].}

\ref\BermanPerryGen{D.S. Berman and M.J. Perry, {\xit ``Generalised
geometry and M-theory''}, \jhep{11}{06}{2011}{074} [\arxiv{1008}{1763}].}    

\ref\BermanMusaevThompson{D.S. Berman, E.T. Musaev and D.C. Thompson,
{\xit ``Duality invariant M-theory: gaugings via Scherk--Schwarz
reduction}, \jhep{12}{10}{2012}{174} [\arxiv{1208}{0020}].}

\ref\UdualityMembranes{V. Bengtsson, M. Cederwall, H. Larsson and
B.E.W. Nilsson, {\xit ``U-duality covariant
membranes''}, \jhep{05}{02}{2005}{020} [\hepth{0406223}].}

\ref\ObersPiolineU{N.A. Obers and B. Pioline, {\xit ``U-duality and M-theory''},
\PR{318}{1999}{113}, 
\nlni [\hepth{9809039}].}

\ref\BermanGodazgarPerry{D.S. Berman, H. Godazgar and M.J. Perry,
{\xit ``SO(5,5) duality in M-theory and generalized geometry''},
\PLB{700}{2011}{65} [\arxiv{1103}{5733}].} 

\ref\BermanMusaevPerry{D.S. Berman, E.T. Musaev and M.J. Perry,
{\xit ``Boundary terms in generalized geometry and doubled field theory''},
\PLB{706}{2011}{228} [\arxiv{1110}{3097}].} 

\ref\BermanGodazgarGodazgarPerry{D.S. Berman, H. Godazgar, M. Godazgar  
and M.J. Perry,
{\xit ``The local symmetries of M-theory and their formulation in
generalised geometry''}, \jhep{12}{01}{2012}{012}
[\arxiv{1110}{3930}].} 

\ref\BermanGodazgarPerryWest{D.S. Berman, H. Godazgar, M.J. Perry and
P. West,
{\xit ``Duality invariant actions and generalised geometry''}, 
\jhep{12}{02}{2012}{108} [\arxiv{1111}{0459}].} 

\ref\CoimbraStricklandWaldram{A. Coimbra, C. Strickland-Constable and
D. Waldram, {\xit ``$E_{d(d)}\times\hbox{\eightbbb R}^+$ generalised geometry,
connections and M theory'' }, \arxiv{1112}{3989}.} 

\ref\CremmerPopeI{E. Cremmer, B. Julia, H. L\"u and C.N. Pope,
{\xit ``Dualisation of dualities. I.''}, \NPB{523}{1998}{73} [\hepth{9710119}].}

\ref\HullT{C.M. Hull, {\xit ``A geometry for non-geometric string
backgrounds''}, \jhep{05}{10}{2005}{065} [\hepth{0406102}].}

\ref\HullM{C.M. Hull, {\xit ``Generalised geometry for M-theory''},
\jhep{07}{07}{2007}{079} [\hepth{0701203}].}

\ref\HullDoubled{C.M. Hull, {\xit ``Doubled geometry and
T-folds''}, \jhep{07}{07}{2007}{080}
[\hepth{0605149}].}

\ref\HullTownsend{C.M. Hull and P.K. Townsend, {\xit ``Unity of
superstring dualities''}, \NPB{438}{1995}{109} [\hepth{9410167}].}

\ref\PalmkvistHierarchy{J. Palmkvist, {\xit ``Tensor hierarchies,
Borcherds algebras and $E_{11}$''}, \jhep{12}{02}{2012}{066}
[\arxiv{1110}{4892}].} 

\ref\deWitNicolaiSamtleben{B. de Wit, H. Nicolai and H. Samtleben,
{\xit ``Gauged supergravities, tensor hierarchies, and M-theory''},
\jhep{02}{08}{2008}{044} [\arxiv{0801}{1294}].}

\ref\deWitSamtleben{B. de Wit and H. Samtleben,
{\xit ``The end of the $p$-form hierarchy''},
\jhep{08}{08}{2008}{015} [\arxiv{0805}{4767}].}

\ref\CederwallJordanMech{M.~Cederwall, {\xit ``Jordan algebra
dynamics''}, \PLB{210}{1988}{169}.} 

\ref\BerkovitsNekrasovCharacter{N. Berkovits and N. Nekrasov, {\xit
    ``The character of pure spinors''}, \LMP{74}{2005}{75}
  [\hepth{0503075}].}

\ref\HitchinLectures{N. Hitchin, {``\xit Lectures on generalized
geometry''}, \arxiv{1010}{2526}.}

\ref\KoepsellNicolaiSamtleben{K. Koepsell, H. Nicolai and
H. Samtleben, {\xit ``On the Yangian $[Y(e_8)]$ quantum symmetry of
maximal supergravity in two dimensions''}, \jhep{99}{04}{1999}{023}
[\hepth{9903111}].}

\ref\HohmHullZwiebachI{O. Hohm, C.M. Hull and B. Zwiebach, {\xit ``Background
independent action for double field
theory''}, \jhep{10}{07}{2010}{016} [\arxiv{1003}{5027}].}

\ref\HohmHullZwiebachII{O. Hohm, C.M. Hull and B. Zwiebach, {\xit
``Generalized metric formulation of double field theory''},
\jhep{10}{08}{2010}{008} [\arxiv{1006}{4823}].} 

\ref\HohmZwiebach{O. Hohm and B. Zwiebach, {\xit ``On the Riemann
tensor in double field theory''}, \jhep{12}{05}{2012}{126}
[\arxiv{1112}{5296}].} 

\ref\WestEEleven{P. West, {\xit ``$E_{11}$ and M theory''},
\CQG{18}{2001}{4443} [\hepth{0104081}].}

\ref\AndriotLarforsLustPatalong{D. Andriot, M. Larfors, D. L\"ust and
P. Patalong, {\xit ``A ten-dimensional action for non-geometric
fluxes''}, \jhep{11}{09}{2011}{134} [\arxiv{1106}{4015}].}

\ref\AndriotHohmLarforsLustPatalongI{D. Andriot, O. Hohm, M. Larfors,
D. L\"ust and 
P. Patalong, {\xit ``A geometric action for non-geometric
fluxes''}, \PRL{108}{2012}{261602} [\arxiv{1202}{3060}].}

\ref\AndriotHohmLarforsLustPatalongII{D. Andriot, O. Hohm, M. Larfors,
D. L\"ust and 
P. Patalong, {\xit ``Non-geometric fluxes in supergravity and double
field theory''}, Fortsch. Phys. {\xbold60} (2012) 1150 [\arxiv{1204}{1979}].}

\ref\DamourHenneauxNicolai{T. Damour, M. Henneaux and H. Nicolai,
{\xit ``Cosmological billiards''}, \CQG{20}{2003}{R145} [\hepth{0212256}].}

\ref\DamourNicolai{T. Damour and H. Nicolai, 
{\xit ``Symmetries, singularities and the de-emergence of space''},
\arxiv{0705}{2643}.}

\ref\EHTP{F. Englert, L. Houart, A. Taormina and P. West,
{\xit ``The symmetry of M theories''},
\jhep{03}{09}{2003}{020}2003 [\hepth{0304206}].}

\ref\PachecoWaldram{P.P. Pacheco and D. Waldram, {\xit ``M-theory,
exceptional generalised geometry and superpotentials''},
\jhep{08}{09}{2008}{123} [\arxiv{0804}{1362}].}

\ref\DamourHenneauxNicolaiII{T. Damour, M. Henneaux and H. Nicolai,
{\xit ``$E_{10}$ and a 'small tension expansion' of M theory''},
\PRL{89}{2002}{221601} [\hepth{0207267}].}

\ref\KleinschmidtNicolai{A. Kleinschmidt and H. Nicolai, {\xit
``$E_{10}$ and $SO(9,9)$ invariant supergravity''},
\jhep{04}{07}{2004}{041} [\hepth{0407101}].}

\ref\WestII{P.C. West, {\xit ``$E_{11}$, $SL(32)$ and central charges''},
\PLB{575}{2003}{333} [\hepth{0307098}].}

\ref\KleinschmidtWest{A. Kleinschmidt and P.C. West, {\xit
``Representations of $G^{+++}$ and the r\^ole of space-time''},
\jhep{04}{02}{2004}{033} [\hepth{0312247}].}

\ref\WestIII{P.C. West, {\xit ``$E_{11}$ origin of brane charges and
U-duality multiplets''}, \jhep{04}{08}{2004}{052} [\hepth{0406150}].}

\ref\PiolineWaldron{B. Pioline and A. Waldron, {\xit ``The automorphic
membrane''}, \jhep{04}{06}{2004}{009} [\hepth{0404018}].}

\ref\WestBPS{P.C. West, {\xit ``Generalised BPS conditions''},
\arxiv{1208}{3397}.}

\ref\BermanCederwallKleinschmidtThompson{D.S. Berman, M. Cederwall,
A. Kleinschmidt and D.C. Thompson, {\xit ``The gauge structure of
generalised diffeomorphisms''}, \jhep{13}{01}{2013}{64} [\arxiv{1208}{5884}].}

\ref\PalmkvistBorcherds{J. Palmkvist, {\xit ``Borcherds and Kac--Moody
  extensions of simple finite-dimensional Lie algebras''}, \arxiv{1203}{5107}.}

\ref\ParkSuh{J.-H. Park and Y. Suh, {\xit ``U-geometry: SL(5)''},
  \arxiv{1302}{1652}.}

\ref\CederwallEdlundKarlsson{M. Cederwall, J. Edlund and A. Karlsson,
  {\xit ``Exceptional geometry and tensor fields''}, \arxiv{1302}{6736}.}

\ref\PalmkvistDual{J. Palmkvist, work in progress.}

\ref\CederwallPalmkvistSerre{M. Cederwall and J. Palmkvist, {\xit
    ``Serre relations, constraints and partition functions''}, to appear.}

\ref\CoimbraStricklandWaldramII{A. Coimbra, C. Strickland-Constable and
D. Waldram, {\xit ``Supergravity as generalised geometry II:
$E_{d(d)}\times\hbox{\eightbbb R}^+$ and M theory'' }, \arxiv{1212}{1586}.}  

\ref\HohmZwiebachGeometry{O. Hohm and B. Zwiebach, {\xit ``Towards an
invariant geometry of double field theory''}, \arxiv{1212}{1736}.} 

\ref\JeonLeeParkI{I. Jeon, K. Lee and J.-H. Park, {\xit ``Differential
geometry with a projection: Application to double field theory''},
\jhep{11}{04}{2011}{014} [\arxiv{1011}{1324}].}

\ref\JeonLeeParkII{I. Jeon, K. Lee and J.-H. Park, {\xit ``Stringy
differential geometry, beyond Riemann''}, 
\PRD{84}{2011}{044022} [\arxiv{1105}{6294}].}

\ref\JeonLeeParkIII{I. Jeon, K. Lee and J.-H. Park, {\xit
``Supersymmetric double field theory: stringy reformulation of supergravity''},
\PRD{85}{2012}{081501} [\arxiv{1112}{0069}].}

\ref\JeonLeeParkRR{I. Jeon, K. Lee and J.-H. Park, {\xit
``Ramond--Ramond cohomology and O(D,D) T-duality''},
\jhep{12}{09}{2012}{079} [\arxiv{1206}{3478}].} 

\ref\HohmKwak{O. Hohm and S.K. Kwak, {\xit ``$N=1$ supersymmetric
double field theory''}, \jhep{12}{03}{2012}{080} [\arxiv{1111}{7293}].}

\ref\HohmKwakFrame{O. Hohm and S.K. Kwak, {\xit ``Frame-like geometry
of double field theory''}, \JPA{44}{2011}{085404} [\arxiv{1011}{4101}].}

\ref\HohmKwakZwiebachI{O. Hohm, S.K. Kwak and B. Zwiebach, {\xit
``Unification of type II strings and T-duality''},
\PRL{107}{2011}{171603} [\arxiv{1106}{5452}].}

\ref\HohmKwakZwiebachII{O. Hohm, S.K. Kwak and B. Zwiebach, {\xit
``Double field theory of type II strings''}, \jhep{11}{09}{2011}{013}
[\arxiv{1107}{0008}].}

\ref\PureSGI{M. Cederwall, {\xit ``Towards a manifestly supersymmetric
    action for D=11 supergravity''}, \jhep{10}{01}{2010}{117}
    [\arxiv{0912}{1814}].}  

\ref\PureSGII{M. Cederwall, 
{\xit ``D=11 supergravity with manifest supersymmetry''},
    \MPLA{25}{2010}{3201} [\arxiv{1001}{0112}].}

\ref\Hillmann{C. Hillmann, {\xit ``Generalized $E_{7(7)}$ coset
dynamics and $D=11$ supergravity''}, \jhep{09}{03}{2009}{135}
[\arxiv{0901}{1581}].}

\ref\HohmZwiebachLarge{O. Hohm and B. Zwiebach, {\xit ``Large gauge
transformations in double field theory''}, \jhep{13}{02}{2013}{075}
[\arxiv{1207}{4198}].} 

\ref\CremmerLuPopeStelle{E. Cremmer, H. L\"u, C.N. Pope and
K.S. Stelle, {\xit ``Spectrum-generating symmetries for BPS solitons''},
\NPB{520}{1998}{132} [\hepth{9707207}].}


\headtext={M. Cederwall: 
``Non-gravitational exceptional supermultiplets''}

\line{
\epsfxsize=18mm
\epsffile{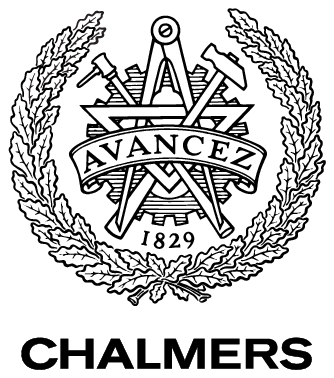}
\hfill}
\vskip-12mm
\line{\hfill Gothenburg preprint}
\line{\hfill February, {\old2013}}
\line{\hrulefill}

\vfill
\vskip.5cm

\centerline{\sixteenhelvbold
Non-gravitational exceptional supermultiplets}

\vfill

\centerline{\twelvehelvbold{Martin Cederwall}}

\vfill

\centerline{\xrm Fundamental Physics}
\centerline{\xrm Chalmers University of Technology}
\centerline{\xrm SE 412 96 Gothenburg, Sweden}

\vfill

{\narrower\noindent \underbar{Abstract:} We examine non-gravitational
  minimal supermultiplets which are based on the tensor gauge fields
  appearing as matter fields in exceptional generalised geometry. When
  possible, off-shell multiplets are given. 
The fields in the multiplets describe non-gravitational parts of the
internal dynamics of 
compactifications of M-theory. 
In flat backgrounds, they enjoy a global U-duality symmetry, but also
provide multiplets with a possibility of coupling to a generalised
exceptional geometry.
\smallskip}
\vfill

\font\xxtt=cmtt6

\vtop{\baselineskip=.6\baselineskip\xxtt
\line{\hrulefill}
\catcode`\@=11
\line{email: martin.cederwall@chalmers.se\hfill}
\catcode`\@=\active
}

\eject

\def\textfrac#1#2{\raise .45ex\hbox{\the\scriptfont0 #1}\nobreak\hskip-1pt/\hskip-1pt\hbox{\the\scriptfont0 #2}}

\def\L{\Lambda}
\def\s{\sigma}
\def\ta{{\tilde a}}

\def\tal{{\tilde\a}}

\def\te{{\tilde\e}}
\def\tchi{{\tilde\chi}}
\def\tH{{\tilde H}}
\def\tphi{{\tilde\phi}}
\def\dee{\d_{\e,\te}}

\def\ra{\rightarrow}


\section\Background{Background}Generalised geometry provides a way of
extending the geometric picture 
of the gravity field to massless tensor fields in string theory or
M-theory, thus manifesting and giving a geometric framework to 
T-duality or U-duality 
[\HullTownsend\skipref\CremmerLuPopeStelle\skipref\CremmerPopeI-\ObersPiolineU].
There has recently been progress in the formulation of generalised
geometric models, both for the doubled field theories (manifesting
T-duality)
[\HitchinLectures\skipref\HullT\skipref\HullDoubled\skipref\HohmHullZwiebachI\skipref\HohmHullZwiebachII\skipref\HohmZwiebach\skipref\HohmZwiebachGeometry\skipref\AndriotLarforsLustPatalong\skipref\AndriotHohmLarforsLustPatalongI\skipref\AndriotHohmLarforsLustPatalongII\skipref\JeonLeeParkI\skipref\JeonLeeParkII\skipref\JeonLeeParkIII\skipref\HohmKwak\skipref\HohmKwakFrame\skipref\HohmKwakZwiebachI\skipref\HohmKwakZwiebachII-\HohmZwiebachLarge] 
and exceptional theories (U-duality) [\HullM\skipref\Hillmann\skipref\BermanPerryGen\skipref\BermanGodazgarPerry\skipref\BermanMusaevThompson\skipref\BermanMusaevPerry\skipref\BermanGodazgarGodazgarPerry\skipref\BermanGodazgarPerryWest\skipref\PachecoWaldram\skipref\CoimbraStricklandWaldram\skipref\CoimbraStricklandWaldramII\skipref\BermanCederwallKleinschmidtThompson\skipref\ParkSuh-\CederwallEdlundKarlsson].
In particular, the tensor gauge fields for the exceptional setting
were described in an accompanying paper,
ref. [\CederwallEdlundKarlsson], together with a tensor calculus for
exceptional generalised geometry.
The purpose of the present letter is to construct supermultiplets, not
containing generalised gravity, based on the known tensor fields.

Bosonic matter fields, 
apart from scalars, should come in the modules
$R_k$, some of which are listed in Table 1. 
These modules play a number of r\^oles in exceptional
geometry
[\BermanCederwallKleinschmidtThompson,\PalmkvistHierarchy\skipref\deWitNicolaiSamtleben-\deWitSamtleben]. In
ref. [\BermanCederwallKleinschmidtThompson], they were 
shown to describe the reducibility of generalised
diffeomorphisms. Their dynamics, as tensor fields, was examined in
ref. [\CederwallEdlundKarlsson]. 
Below, we will realise minimal 
global supersymmetry in $n=4,5,6$ on matter multiplets.
As will be clear, ``$N=1$'' supersymmetry in $n=4,5,6,7$ contains 
4, 8, 16 and 32 supersymmetries, respectively. The real forms may need
to be chosen differently in order to have real fields.
We will constrain ourselves to tensor potentials in the modules
$R_2,\ldots,R_{8-n}$. There may also be fields in $R_1$, which is the
``generalised graviphoton'', but its number of degrees of freedom is
too large to allow for minimal set of fermions only.

The sequences $\{R_k\}$ are infinite, but relevant tensor fields come
in $R_1,\ldots,R_{8-n}$, which is the part of the sequence where
derivatives from $R_k$ to $R_{k-1}$ is connection-free. These modules
are analogous to forms, both in the absence of connection, and in the
nilpotency of the derivative, which makes the concepts of field
strengths and gauge transformations natural. Details are found in
ref. [\CederwallEdlundKarlsson].

\TightTables
\vskip4\parskip
\ruledtable
$\,n$ | $G$ | $\ol H$ | $R_1$ & $R_2$ & $R_3$ & $R_4$ & $R_5$ \crthick
$\,3$ | $SL(3)\times SL(2)$ | $SU(2)\times U(1)$ | $({\bf3},{\bf2})$ &
 $(\overline{\bf3},{\bf1})$ & $({\bf1},{\bf2})$ & $({\bf3},{\bf1})$ &
 $(\overline{\bf3},\overline{\bf2})$ \cr  
$\,4$ | $SL(5)$ | $Spin(5)$ | ${\bf10}$ & $\overline{\bf5}$ & ${\bf5}$ &
 $\overline{\bf10}$ & ${\bf24}$ \cr   
$\,5$ | $SO(5,5)$ | $Spin(5)\times Spin(5)$ | $\bf16$ & $\bf10$ &
 $\overline{\bf16}$ & $\bf45$ \cr  
$\,6$ | $E_{6(6)}$ | $USp(8)$ | $\bf27$ & $\overline{\bf27}$ & $\bf78$ &
 \cr
$\,7$ | $E_{7(7)}$ | $SU(8)$ | $\bf{56}$ & $\bf{133}$ 
 \endruledtable
\Table\ReducibilityTable{A partial list of modules $R^{(n)}_k$.}
\LooseTables

\section\Construction{Construction of the multiplets}As a guide for
the construction, it is informative to list the 
counting of off-shell and on-shell degrees of freedom in the tensor
fields. In the cases we consider, the number of off-shell and on-shell
degrees of freedom
of $R_k$ are given by the dimensions of
the corresponding modules when $n$ is lowered by one and two units, 
respectively [\CederwallEdlundKarlsson].

\vskip4\parskip
\ruledtable
 $n$ | $R_2$ | $R_3$ | $R_4$ \crthick
 $4$ | $3\ra2$ | $2\ra1$ | $3\ra1$ \cr
 $5$ | $5\ra3$ | $5\ra2$ | \cr
 $6$ | $10\ra5$ ||
\endruledtable
\Table\CountingTable{Off-shell and on-shell counting for $R_k$.}

The fermions (``spinor'' fields and supersymmetry generators) 
transform in a module $S$ of the double cover $\ol H$ of the compact
subgroup of the U-duality group. For $n=4$ the spinor module is
$S={\bf4}=(01)$ of $Spin(5)$, for $n=5$
$S=({\bf4},{\bf1})\oplus({\bf1},{\bf4})=(01)(00)\oplus(00)(01)$ of
$Spin(5)\times Spin(5)$, for $n=6$ $S={\bf8}=(1000)$ of $USp(8)$, and
for $n=7$ $S={\bf8}\oplus\ol{\bf8}=(1000000)\oplus(0000001)$ of $SU(8)$. 
These modules may also come together with some R-symmetry. In fact,
some non-trivial R-symmetry must be present in $n=6,7$. This is
because the momentum module ${\bf27}=(0100)$ or
${\bf28}\oplus\ol{\bf28}=(0100000)\oplus(0000010)$ only comes in the  
antisymmetric product $\wedge^2S$. So in those cases, there should be
at least an $SU(2)$ or $SL(2)$ R-symmetry providing a two-index $\e$
tensor. What we here call 
R-symmetry is of course a Lorentz symmetry, from the usual perspective
of compactification.
It is not possible to have ``chiral'' spinors in $n=5$, since the
momenta are in $\ol{\bf16}\rightarrow({\bf4},{\bf4})$.

This means that minimal supersymmetry implies 
4, 8, 16 and 32 supersymmetries for $n=4$, 5, 6 and 7, respectively.
There may also be restrictions connected to the real form of the
U-duality group and the local subgroup. For $n=4$ and compact
$Spin(5)$, $S={\bf4}$ is not real. 
We will nevertheless construct the minimal supermultiplets based on
$R_2$, $R_3$ and $R_4$ for $n=4$, on $R_2$ and $R_3$ for $n=5$ and
on $R_2$ for $n=6$. For $n=7$, minimal supersymmetry will imply 32
supercharges, unless one may build a model with ``chiral'' spinors.

We always assume flat backgrounds. The coupling to some non-trivial
generalised geometry should be straightforward, along the lines of
ref. [\CederwallEdlundKarlsson]. The derivatives used to construct
field strengths from gauge potentials and equations of motion from
dual field strengths are free of connection. 

\vfill\eject

\subsection\NFourRTwo{$n=4$, $R_2$}We take the bosonic field to be a potential 
$A_m$ in $R_2=\bf{\ol5}$ of $SL(5)$ and 
the fermion to be a spinor $\chi^\a$ in $\bf4$ of $Spin(5)$. In view
of Table \CountingTable, this should be enough to match the 2 on-shell
degrees of freedom of a spinor. One auxiliary scalar should help
supersymmetry close off-shell.
The potential has a field strength in $R_1=\bf10$ of $G$:
$F_{mnp}=3\*_{[mn}A_{p]}$. The gauge transformations with $\L$ in
$R_3={\bf\ol5}$ are 
$\d_\L A_m=\*_{mn}\L^n$, and invariance of $F$ relies on the section
condition $\*_{[mn}\*_{pq]}=0$.

The supersymmetry transformations are
$$
\eqalign{
\d_\e A_m&=(\e\g_m\chi)\komma\cr
\d_\e\chi^\a&=\fr6F_{mnp}(\g^{mnp}\e)^\a\punkt\cr
}\eqn
$$
We will check the supersymmetry algebra. The general symmetric 
Fierz identity is
$A^{(\a}B^{\b)}=-\fr8(\g_{ab})^{\a\b}(A\g^{ab}B)$,
which simplifies Fierz rearrangements.
Commuting two supersymmetry transformations on $A_m$ gives
$$
\eqalign{
[\d_{\e'},\d_\e]A_m&=\fr6(\e\g_m\g^{npq}\e')F_{npq}-(\e\leftrightarrow\e')\cr
&=(\e\g^{np}\e')\*_{np}A_m+2\*_{mn}\left[(\e\g^{np}\e')A_p\right]\komma\cr
}\eqn
$$
\ie, a translation and a gauge transformation.

Acting on the fermion, one gets
$$
\eqalign{
[\d_{\e'},\d_\e]\chi^\a&=\fr2(\g^{mnp}\e)^\a(\e'\g_m\*_{np}\chi)
                 -(\e\leftrightarrow\e')\cr
&=-\fr8(\e\g^{rs}\e')(\g^{mnp}\g_{rs}\g_m\*_{np}\chi)^\a\cr
&=(\e\g^{np}\e')\*_{np}\chi^\a
       +\fr8(\e\g^{rs}\e')(\g_{rs}\g^{np}\*_{np}\chi)^\a\punkt\cr
}\eqn
$$
In the last step, we have used the identity
$\g^{mnp}\g_{rs}\g_m+\g_{rs}\g^{np}=-8\d^{np}_{rs}$, which happens to
be valid in five dimensions. The first term on the last line 
is the translation, with
the same coefficient as on $A$, and the second one an equation of
motion. 

Taking the supersymmetry variation on the fermion equation of motion
should give the one for the bosons.
$$
\eqalign{
&\d_\e(\g^{mn}\*_{mn}\chi)^\a=\fr6(\g^{mn}\g^{pqr}\e)^\a\*_{mn}F_{pqr}\cr
&\qquad=-(\g^m\e)^\a\*^{np}F_{mnp}-(\g^{npq}\e)^\a\*^m{}_nF_{mpq}
                 +\fr6(\g^{mnpqr}\e)^\a\*_{mn}F_{pqr}\cr
&\qquad=-(\g^m\e)^\a\*^{np}F_{mnp}-(\g^{npq}\e)^\a\*_{[mn}\*_{pq]}A^m
             +\fr6(\g^{mnpqr}\e)^\a\*_{[mn}\*_{pq]}A_r\punkt\cr
}\Eqn\OnShellA
$$
So, no ``extra'' conditions are produced, except for the equation of
motion $\*^{np}F_{mnp}=0$, if the section condition (the last two
terms) is fulfilled.

One can indeed introduce a single scalar auxiliary field $H$, with the
supersymmetry transformations
$$
\eqalign{
\d_\e A_m&=(\e\g_m\chi)\komma\cr
\d_\e\chi^\a&=\fr6F_{mnp}(\g^{mnp}\e)^\a+H\e^\a\komma\cr
\d_\e H&=\fr2(\e\g^{mn}\*_{mn}\chi)\punkt\cr
}\eqn
$$
This cancels the equation of motion term in eq. (\OnShellA), does not
affect the commutator acting on $A_m$ and gives the correct algebra on
$H$. The closure on $H$ is provided by
$$
\eqalign{
&[\d_{\e'},\d_\e]H=\fr2(\e\g^{mn}\e')\*_{mn}H
   +\fr{12}(\e\g^{mn}\g^{pqr}\e')\*_{mn}F_{pqr}-(\e\leftrightarrow\e')\cr
&\qquad=(\e\g^{mn}\e')\*_{mn}H+(\e\g^{mpq}\e')\*_m{}^nF_{pqn}\punkt\cr
}\eqn
$$
The expression $\*_{[m}{}^nF_{pq]n}$ vanishes due to the section condition.
The equations of motion follow by supersymmetry from $H=0$.

\subsection\NFourRThree{$n=4$, $R_3$}Consider a potential $B^m$ 
in $R_3={\bf5}$, 
with field strength $G_m=\*_{mn}B^n$ in $R_2={\bf\ol5}$ and
gauge transformation $\d_\L B^m=\*_{np}\L^{[mnp]}$ with $\L$ in
$R_4={\bf\overline{10}}$. In addition to a
fermion $\chi^\a$, one needs a
physical scalar field $\phi$, and for off-shell supersymmetry an auxiliary
field $H$. The transformations are
$$
\eqalign{
\d_\e B^m&=(\e\g^m\chi)\komma\cr
\d_\e \phi&=(\e\chi)\komma\cr
\d_\e \chi^\a&=-G_m(\g^m\e)^\a+\fr2\*_{mn}\phi(\g^{mn}\e)^\a
                       +H\e^\a\komma\cr
\d_\e H&=\fr2(\e\g^{mn}\*_{mn}\chi)\punkt\cr
}\eqn
$$
The commutator on $\chi^\a$ gives
$$
\eqalign{
[\d_{\e'},\d_\e]\chi^\a&=(\e\g^{pq}\e')
\left((\fr4\g^m\g_{pq}\g^n-\fr8\g^{mn}\g_{pq}-\fr8\g_{pq}\g^{mn})
             \*_{mn}\chi\right)^\a\cr
&\qquad=(\e\g^{mn}\e')\*_{mn}\chi^\a\komma\cr
}\eqn
$$
and acting on $B^m$ one gets
$$
\eqalign{
&[\d_{\e'},\d_\e]B^m=-(\e\g^m\g^n\e')\*_{np}B^p
         +\fr2(\e\g^m\g^{np}\e')\*_{np}\phi+(\e\g^m\e')H
         -(\e\leftrightarrow\e')\cr
&\qquad=(\e\g^{np}\e')\*_{np}B^m
   +\*_{np}\left[-3(\e\g^{[mn}\e')B^{p]}+(\e\g^{mnp}\e')\phi\right]\punkt\cr
}\eqn
$$
The commutators on $\phi$ and $H$ also work, the latter thanks
to the section condition, giving the Bianchi identity $\*_{[mn}G_{p]}=0$.

All equations of motion for the physical fields 
are generated by supersymmetry from $H=0$, and
are, as expected:
$$
\eqalign{
\*_{mn}G^n&=0\komma\cr
\*^{mn}\*_{mn}\phi&=0\komma\cr
(\g^{mn}\*_{mn}\chi)^\a&=0\punkt\cr
}\eqn
$$

\subsection\NFourRFour{$n=4$, $R_4$}The last example for $n=4$ is a
potential $C_{mn}$ in 
$R_4={\bf\overline{10}}$ with a field strength
$H_{mnpq}=6\*_{[mn}C_{pq]}$ in $R_3={\bf5}$. Gauge transformations
leaving $H$ invariant are $\d_\L C_{mn}=\*_{[m|p|}\L_{n]}{}^p$, with
$\L$ in ${\bf1}\oplus{\bf24}$. 
We also know that this action of a derivative
will contain connection, which is no problem in flat space, but still
somewhat confusing (see ref. [\CederwallEdlundKarlsson] for
comments). 
Until this is cleared out, a covariant formulation of gauge
transformations in a non-trivial generalised gravity background
may be problematic. A scalar $\phi$ is also needed.

Let the fields transform according to
$$
\eqalign{
\d_\e C_{mn}&=(\e\g_{mn}\chi)\komma\cr
\d_\e\phi&=(\e\chi)\komma\cr
\d_\e\chi^\a&=-\fr{24}H_{mnpq}(\g^{mnpq}\e)^\a
            +\fr2\*_{mn}\phi(\g^{mn}\e)^\a\punkt\cr
}\eqn
$$  
This gives a commutator of supersymmetries on $\chi$:
$$
\eqalign{
[\d_{\e'},\d_\e]\chi^\a&=(\e\g^{pq}\e')
\left((\fr{16}\g^{mnrs}\g_{pq}\g_{rs}-\fr8\g^{mn}\g_{pq})
             \*_{mn}\chi\right)^\a\cr
&\qquad=(\e\g^{mn}\e')\*_{mn}\chi^\a\komma\cr
}\eqn
$$
This is an off-shell multiplet, without the inclusion of auxiliary
fields. It is straightforwardly checked that the commutators on the
bosonic fields work out, \eg:
$$
\eqalign{
[\d_{\e'},\d_\e]C_{mn}&=(\e\g^{pq}\e')\*_{pq}C_{mn}\cr
&+\*_{[m|p|}\left[\d_{n]}{}^p(\e\g^{qr}\e')C_{qr}-4(\e\g^{pq}\e')C_{n]q}
      -4(\e\g_{n]}{}^p\e')\phi\right]\punkt\cr
}
\eqn
$$

Assuming the equation of motion for $\chi$ to be the same as above, we
can apply a supersymmetry transformation, and get the equation of
motion for $C$, $\*^{pq}H_{mnpq}=0$, as expected.

\subsection\NFiveRTwo{$n=5$, $R_2$}In $n=5$, consider a potential
$A_m$ in $R_2={\bf10}$ with a field strength 
$F^A$ in $R_1={\bf16}$ and gauge parameter in
$R_3={\bf\overline{16}}$.
The potential splits into $({\bf5},{\bf1})\oplus({\bf1},{\bf5})$ 
under $H=SO(5)\times
SO(5)$, while ${\bf16},{\bf\overline{16}}\rightarrow({\bf4},{\bf4})$.
The supersymmetry parameters $\e,\te$ are in 
$S=({\bf4},{\bf1})\oplus({\bf1},{\bf4})$.
In addition to the gauge potential, there will be spinors $\chi,\tchi$
in $S$ and a scalar $\phi$. We know that $R_2$ has 3 degrees of freedom
on-shell, which together with the scalar 
matches the $\fr2\times8=4$ fermions. Off-shell supersymmetry demands
2 auxiliary fields $H.\tH$.

The supersymmetry transformations are
$$
\eqalign{
&\d_{\e,\te} A_a=(\e\s_a\chi)\komma\qquad\d_{\e,\te}
A_\ta=(\te\s_\ta\tchi)\komma\cr
&\d_{\e,\te}\phi=(\e\chi)+(\te\tchi)\komma\cr
&\d_{\e,\te}\chi^\a=((F+\*\phi)\te)^\a+\e^\a H\komma
\qquad\d_{\e,\te}\tchi^\tal=((-F^t+\*^t\phi)\e)^\tal+\te^\tal\tH\cr
&\d_{\e,\te}H=2(\te\*^t\chi)\komma
   \qquad\d_{\e,\te}\tH=2(\e\*\tchi)\komma\cr
}\eqn
$$
where $F$ and $\*$ are matrices in $({\bf4},{\bf4})$ and
$F^A=(\g^m\*)^AA_m$ gives $F_{\a\tal}=(\s^a\*)_{\a\tal}A_a-(\*\s^\ta)_{\a\tal}A_\ta$.
It is straightforward to derive the commutation relations
$$
[\d_{\e,\te},\d_{\e',\te{}'}]=2\xi^A\*_A+\ldots
$$
where the translation parameter is $\xi=\e\otimes\te{}'-\e'\otimes\te$
and the ellipsis denotes a gauge transformation 
$\d_\L A_m=(\*\g_m\L)$
with
parameter\foot{This expression is not $SO(10)$ 
covariant, but there will be a metric converting ${\bf16}$ to 
${\bf\overline{16}}$.}
$\,\L_A=-(\g^m\xi)_AA_m-\xi_A\phi$.

\subsection\NFiveRTwo{$n=5$, $R_3$}There should also be a multiplet
with potential in  
$R_3={\bf\overline{16}}$ and field strength in $R_2={\bf10}$. 
The number of off-shell degrees of freedom in $R_3$ is 5, and reduces
on-shell to 2. The
multiplet turns out to contain two scalars, and can be taken off-shell
by the introduction of a single auxiliary field.
The transformations are
\multi
$$
\eqalignno{
&\dee A=\e\otimes\tchi+\chi\otimes\te\komma\cr
&\dee\phi=2(\e\chi)\komma\qquad\dee\tphi=2(\te\tchi)\komma
                \cr
&\dee\chi^\a=F_a(\s^a\e)^\a+(\*\phi\te)^\a+H\e^\a\komma
                &\multieqn{}\cr
&\dee\tchi^\tal=F_\ta(\s^\ta\te)^\tal
         +(\*^t\tphi\e)^\tal+H\te^\tal\komma\cr
&\dee H=(\e\*\tchi)+(\te\*^t\chi)\punkt\cr
}
$$

\subsection\NFiveRTwo{$n=6$, $R_2$}For $n=6$, we only consider the
$R_2$ module. One has a  
potential $A$ in $R_2={\bf\overline{27}}$ with a field strength in 
$R_1={\bf27}$. When $E_6\rightarrow USp(8)$,
${\bf27}\rightarrow{\bf27}$,
${\bf\overline{27}}\rightarrow{\bf27}$, which is an $\e$-traceless
antisymmetric tensor. As mentioned in section \Background, at least an
$SU(2)$ R-symmetry is needed, and the fermions come in $({\bf8},{\bf2})$ under 
$USp(8)\times SU(2)$. 
The number of off-shell degrees of
freedom in $R_2$ is 10, and on-shell 5 (it contains effectively a
six-dimensional 1-form and a 4-form, dual to a scalar). It seems
reasonable to expect that $R_2$ is accompanied by an $SU(2)$ triplet
of scalars, so that the 8 on-shell bosonic degrees of freedom match
the fermionic ones. This would correspond to the degrees of freedom of
$N=(1,1)$ SYM in six dimensions, where the $SO(4)$ R-symmetry is
broken to $SU(2)$ by the dualisation of one scalar.

The transformations are
$$
\eqalign{
\d_\e A_{ab}&=4\e_{ij}\e_{[a}^i\chi_{b]'}^j\komma\cr
\d_\e\chi_a^i&=\e^{bc}\left(F_{ab}\e_c^i
    +\e_{jk}\*_{ab}\phi^{ij}\e_c^k\right)\komma\cr
\d_\e\phi^{ij}&=2\e^{ab}\e_a^{(i}\chi_b^{j)}\komma\cr
}\eqn
$$
where $F_{ab}=2\e^{cd}\*_{[a|c|}A_{b]'d}$ and where $X_{[ab]'}=X_{[ab]}-\fr8\e_{ab}\e^{cd}X_{cd}$ denotes $\e$-traceless
antisymmetrisation. The supersymmetry transformations
commute to
$$
[\d_{\e'},\d_\e]=-\e_{ij}\e^{ab}\e^{cd}\e_a^i\e'{}_c^j\*_{bd}\punkt\eqn
$$
modulo gauge transformations and equations of motion.

\section\Summary{Summary and outlook}We have constructed a number of
non-gravitational supermultiplets, which in the present formulation
enjoy global U-duality. Coupling to a generalised geometric background
will be straightforward, along the lines of ref. [\CederwallEdlundKarlsson].

Let us analyse the effective physical content of the multiplets. The
bosonic fields are listed in Table 3.
The scalars appearing in addition to the fields from $R_k$ are given
in brackets. Fields without local degrees of freedom 
($(n-1)$-forms and $n$-forms) are
not listed. Notice how dualisation of some fields makes manifest
U-duality possible. Take for example the $n=6$ multiplet. The physical
content agrees with $N=(1,1)$ super-Yang--Mills theory in six
dimensions, which has an $SU(2)\times SU(2)$ R-symmetry, with the
scalars transforming as $({\bf2},{\bf2})$. By only
considering the diagonal subgroup,
$({\bf2},{\bf2})\ra{\bf1}\oplus{\bf3}$, the singlet can be dualised
into a 4-form, which together with the vector (and an unphysical
vector) build up the $R_2$ module. 

\TightTables
\vtop{
\vskip\parskip
\ruledtable
$\,n$ | $R_2$ | $R_3$ | $R_4$ \crthick
$\,4$ | vector | scalar + [ scalar ] | 2-form + [ scalar ] \cr
$\,5$ | vector + [ scalar ] | scalar + 3-form + [ 2 scalars ] |\cr
$\,6$ | vector + 4-form + [ 3 scalars ] ||
\endruledtable
\Table\ContentTable{The effective content of physical bosonic fields
  in the multiplets.}
}

As mentioned in Section \Construction, we have not put emphasis on
the real forms of the U-duality groups and their locally realised
subgroups. The reality properties of the fields in the multiplets will
depend on the choice of real form.

All the multiples we have described are minimal. Extended
non-gravitational supermultiplets may be obtained by dimensional
reduction from higher to lower $n$. The general branching pattern for
the modules under consideration here is $R^{(n)}_k\rightarrow
R^{(n-1)}_k\oplus R^{(n-1)}_{k+1}$. We have not touched on the issue
of multiplets containing generalised gravity. The minimal exceptional
gravity multiplets were described in
ref. [\CoimbraStricklandWaldramII], and the content of the
maximal supermultiplets was sketched in
ref. [\CederwallEdlundKarlsson].
A detailed investigation, including a superspace formulation would be
interesting, and might lead towards a formalism, generalising that of
refs. [\PureSGI,\PureSGII], 
where U-duality and supersymmetry are simultaneously manifested.

\acknowledgements The author would like to thank David Berman, Malcolm
Perry, Anna Karlsson, Joakim Edlund and M\aa ns Henningson for
discussions and comments.

\refout

\end